\documentclass[12pt,letterpaper]{article}
\pdfoutput=1
\usepackage{jheppub}
\usepackage{amsfonts, amsthm}
\usepackage[english]{babel}
\usepackage[utf8]{inputenc}
\usepackage{slashed}
\usepackage{mathrsfs}
\usepackage{amssymb}
\usepackage{color}
\hypersetup{unicode}
\usepackage{natbib}
\usepackage{xcolor}
\usepackage{amsmath,amscd}
\usepackage{graphicx,shortvrb}
\usepackage{epsfig}
\usepackage{newlfont}
\usepackage{hyperref}

\newcommand{\eq}{\begin{equation}}
\newcommand{\feq}{\end{equation}}
\newcommand{\eqn}{\begin{eqnarray}}
\newcommand{\feqn}{\end{eqnarray}}

\newcommand{\ma}[1]{\mbox{$\mathcal{#1}$}}

\newcommand{\mrm}[1]{\mbox{$\mathrm{#1}$}}

\def\dil{\varphi}
\def\threeformsix{H}

\def\detT{\det \Tmat}
\def\Tmat{T}

\def\poped{\Delta}
\def\popeg{g_0}
\def\popem{\mu}

\def\gprime{k_0}
\def\volthree{{\mathrm{vol}_3}}

\def\isixa{\mu}
\def\isixb{\nu}
\def\isixc{\rho}

\def\ifoura{i}
\def\ifourb{j}
\def\ifourc{k}
\def\ifourd{l}
\def\ifoure{m}
\def\ifourf{n}

\usepackage[normalem]{ulem}

\title{Supersymmetric Dyonic Strings in 6-Dimensions from 3-Dimensions}

\author[a]{Nihat Sadik Deger,}
\author[a]{Nicol\`o Petri}
\author[b,c]{and Dieter Van den Bleeken}

\affiliation[a]{Department of Mathematics, Bo\u{g}azi\c{c}i University,
34342, Bebek, Istanbul, Turkey.}
\affiliation[b]{Primary address: Dept. of Physics, Bo\u{g}azi\c{c}i University,
34342, Bebek, Istanbul, Turkey.}
\affiliation[c]{Secondary address: Institute for Theoretical Physics, KU Leuven
3001 Leuven, Belgium}

\emailAdd{sadik.deger@boun.edu.tr}
\emailAdd{nicolo.petri@boun.edu.tr}
\emailAdd{dieter.van@boun.edu.tr}

\abstract{It was shown in \href{https://arxiv.org/abs/1410.7168}{1410.7168} that compactifying $D=6$, $\mathcal{N}$=(1,0) ungauged supergravity coupled to a single tensor multiplet on S$^3$ one gets a particular $D=3$, $\mathcal{N}$=4 gauged supergravity which is a consistent reduction. We construct two supersymmetric black string solutions in this 3-dimensional model with one and two active scalars respectively. Uplifting the first, one gets a dyonic string solution in $D=6$ that has been known for a long time. Whereas, uplifting 
the second solution, one finds a very interesting configuration where magnetic strings are located uniformly on a circle 
in a plane within the 4-dimensional flat transverse space and electric strings are distributed homogeneously inside this circle. 
Both solutions have $\mathrm{AdS}_3 \times$ S$^3$ limits.}

\begin{document}
\maketitle
\flushbottom

\section{Introduction}
Six dimensional supergravity models have been an active area of research for a long time. Among them the simplest one is
the so called minimal model whose bosonic field content is just a graviton and a two-form field with a self-dual field strength. Because of the self-duality it has no action. Coupling this model with a single tensor multiplet which has a 
dilaton and two-form field with an anti-self dual field strength as bosonic fields,  these two-forms can be combined to obtain one with an unrestricted field strength and we have the following bosonic Lagrangian \cite{Nishino:1986dc}: 
\begin{equation}\label{Lag6}
\mathscr L_6=\sqrt{-g}\Big(R-\frac{1}{2}\partial_\isixa\dil\partial^\isixa\dil-\frac{1}{12}e^{-\sqrt{2}\dil}\threeformsix_{\isixa\isixb\isixc}\threeformsix^{\isixa\isixb\isixc}\Big)
\;.
\end{equation}
This theory can be obtained from Heterotic or type IIB theory on $K3$ or $T^4$ with some truncation. Therefore,
solutions of this model can be embedded to 10-dimensions as well which provides additional motivation for studying them. The general form of supersymmetric solutions of this model and its generalizations with couplings 
of other multiplets have been studied in \cite{Gutowski:2003rg, Cariglia:2004kk, Akyol:2010iz, Cano:2018wnq, Lam:2018jln}.
A big motivation of studying such solutions is to understand microstate geometries of 5-dimensional 
black holes \cite{Bena:2011dd, Niehoff:2012wu, Bobev:2012af, Vasilakis:2013tjs, Niehoff:2013kia, Shigemori:2013lta}. 
Of course, such configurations are also crucial in studying the $\mathrm{AdS}_3/\mathrm{CFT}_2$ correspondence in detail \cite{Lunin:2002bj, 
Lunin:2002iz, Lin:2004nb, Martelli:2004xq, Liu:2004ru, Boni:2005sf}. In particular, the 6d model \eqref{Lag6} admits a 1/4 supersymmetric 
dyonic string solution that carries electric and magnetic charges \cite{Duff:1995yh, Duff:1996cf} which corresponds to the D1-D5 intersection in type IIB theory and upon dimensional reduction on a circle leads to a black hole in $D=5$. Hence, finding new dyonic string solutions is of considerable interest. Known examples include \cite{Duff:1995yh, Duff:1996cf, Duff:1998cr, Gueven:2003uw, RandjbarDaemi:2004qr, Jong:2006za}.

A few years ago it was shown that dimensional reduction of the 6-dimensional model \eqref{Lag6} on S$^3$ leads to
a $\ma N=4, SO(4)$ gauged supergravity \cite{Deger:2014ofa}. Moreover, this is a consistent reduction which means that any solution in the 3-dimensional theory is automatically a solution in 6-dimensions. Such consistent sphere reductions are quite rare and when available they can be used to construct complicated solutions in the higher dimensional theory, which is the main theme of this paper. An SU(2) group manifold reduction of \eqref{Lag6} to 3d is also known and is consistent by construction 
\cite{Gava:2010vz}.

In three dimensions it is possible to formulate supergravities in two different ways
with vector fields appearing in the Yang-Mills (YM) form or the
Chern-Simons (CS) form in the action respectively \cite{Nicolai:2003bp}. They are equivalent to each other and one can go from the CS to YM formulation through some differential constaints. The general construction of 3-dimensional gauged supergravities was given in \cite{deWit:2003ja, deWit:2004yr} using the CS formulation. Yet, the model one obtains 
from a dimensional reduction is in YM form. In  \cite{Deger:2014ofa} the CS form of 
the aforementioned $\ma N=4, SO(4)$ theory was also identified. Supersymmetry transformations (or BPS conditions that follow from them by setting fermions to zero) are given in \cite{deWit:2003ja, deWit:2004yr} and can be carried to YM formulation by the help of the duality constraint equations as we do in this paper.

In the next section we begin with a brief description of our 3-dimensional model in the YM form. It contains 10 scalars and 6 vectors which makes it hard to search for exact solutions. We first simplify this theory by truncating it to a subsector invariant under the $U(1) \times U(1)$ subgroup of the $SO(4)$ gauge group after which only 2 vectors and 4 scalars remain. In section 3, using the BPS conditions for this sector  
we construct two different uncharged black string solutions with one and two active scalars respectively. Then, in section 4 we uplift these to $D=6$. The first one leads to a well-known dyonic string solution that was found long ago \cite{Duff:1995yh} describing a single electric and single magnetic charge located at the origin of the 4-dimensional flat transverse space. The uplift of the latter,
however, results in a rather peculiar configuration where magnetic strings are located uniformly on a circle in a plane in the 4-dimensional transverse space and electric strings are distributed homogeneously inside this circle. Both of these solutions have $\mathrm{AdS}_3 \times$ S$^3$ limits. We conclude with some remarks and future directions in section 5. Derivation of the BPS conditions is given in appendix \ref{newapp}.

\section{3-Dimensional $\ma N=4$, $SO(4)$  Gauged Supergravity}
\label{section:N4sugra}
The 3d supergravity model that we are interested in can be obtained from $\ma N=(1,0)$ 6d ungauged supergravity coupled to a single tensor multiplet by a consistent 3-sphere reduction \cite{Deger:2014ofa}. This theory preserves 8 real supercharges, i.e. $\ma N=4$, and its bosonic Lagrangian is \cite{Deger:2014ofa}:
 \begin{equation}
\begin{split}
 \sqrt{-g}^{\,-1}\mathscr{L}&=R-\frac14 T^{-1}_{ij}T^{-1}_{kl}D_\mu T_{jk}D^{\mu}T_{li}-
 \frac18 T^{-1}_{ik}T^{-1}_{jl} F_{\mu \nu\, ij}F^{\mu \nu}_{kl}-V\\
 &-\frac{k_0}{8}\,\sqrt{-g}^{\,-1}\,\epsilon_{ijkl}\,\varepsilon^{\mu\nu\rho}A_{\mu\, ij}\left(\partial_\nu A_{\rho\, kl}+\frac23 g_0 
 A_{\nu\, km}A_{\rho\, ml}\right)\,,
 \label{so4action}
 \end{split}
\end{equation}
with $i,\,j,\,k=1,\dots,4$. The theory \eqref{so4action} is manifestly $SO(4)$ covariant and it depends explicitly on the symmetric matrix $T_{ij}$ parametrizing the quaternionic target manifold
 \begin{equation}
 \frac{GL(4)}{SO(4)}\subset\frac{SO(4,4)}{SO(4)\times SO(4)}\,.
 \label{sigmamodel}
\end{equation}
Its gauge group $SO(4)$ determines the following scalar potential
\begin{equation}
 V=\frac12 \left(k_0^2 \det T+2g_0^2 T_{ij}T_{ij}-g_0^2 (T_{ii})^2 \right)\,.
 \label{scalarpotential}
\end{equation}
The covariant derivatives and the field strengths are respectively given by
\begin{equation}
\begin{split}
 &D_\mu T_{ij}=\partial_\mu T_{ij}+ g_0 A_{\mu\,ik}T_{kj}+g_0 A_{\mu\,jk}T_{ki}\,, \\
 &F_{\mu\nu\,ij}= 2\partial_{[\mu} A_{\nu]\,ij}+g_0 A_{\mu \,ik} A_{\nu\,kj}-g_0 A_{\mu \,jk} A_{\nu\,ki}\,.
 \end{split}
\end{equation}
To proceed, we simplify the theory by considering a further truncation, that is consistent by symmetry considerations. The particular symmetry we choose to preserve is
\begin{equation}
 U(1)\times U(1) \subset SO(3)\times SO(3)\simeq SO(4) \, .
 \label{so4break}
\end{equation}
The matrix $T_{ij}$ is taken to depend only on the four real scalar fields 
$\phi^i=(\xi_1,\xi_2,\rho,\theta)$ and it is of the following block diagonal form:
\begin{equation}
T=\left(
 \begin{array}{cc}
  e^{\xi_1}e^{R(\rho,\theta)} \mathbb{I}_2 & 0_2\\
  0_2 & e^{\xi_2}\,\mathbb{I}_2
 \end{array} \right)\qquad \text{with} \qquad R(\rho,\theta)=
\rho\, \left(
 \begin{array}{cc}
  \sin\theta & \cos\theta\\
  \cos\theta & -\sin\theta
 \end{array} \right)\,.
 \label{parT}
\end{equation}
The vectors $A_{\mu\,ij}$ respecting \eqref{so4break} have the form 
\begin{equation}
 A_\mu=
 \left(
 \begin{array}{cc}
   A^1_\mu & 0\\
  0 &  A^2_\mu
 \end{array} \right)
 \qquad \text{with} \qquad A_\mu^{1,2}=
\left(
 \begin{array}{cc}
  0 & \ma A_\mu^{1,2}\\
  -\ma A_\mu^{1,2} & 0
 \end{array} \right)\, ,
 \label{parA}
\end{equation}
where $\ma A_\mu^{1,2}$ are two abelian vector fields.
If we express the YM Lagrangian \eqref{so4action} in this explicit parametrization, we obtain\footnote{
The abelian Chern-Simons term was inadvertently missing in earlier versions of this paper, which was noticed 
after comparison with \cite{Mayerson:2020tcl}. This extra term does not affect the analysis in the remainder of this paper.
}
 \begin{equation}
\begin{split}
 \sqrt{-g}^{\,-1}\mathscr{L}&=R-\frac12 (\partial_\mu\,\xi_1)^2-\frac12 (\partial_\mu\,\xi_2)^2 -\frac12 (\partial_\mu\,\rho)^2-\frac12\,\sinh^2\rho (D_\mu\,\theta)^2\\
 &-\frac14\,e^{-2\xi_1}\,\ma F^1_{\mu\nu}\ma F^{1\,\mu\nu}-\frac14\,e^{-2\xi_2}\,\ma F^2_{\mu\nu}\ma F^{2\,\mu\nu}
-\frac{k_0}{2}\,\sqrt{-g}^{\,-1}\,\varepsilon^{\mu\nu\rho} \ma A_\mu^{1} \ma F^2_{\nu\rho}
 -V\,,
 \end{split}
 \label{actionN=4model}
\end{equation}
with the covariant derivative
\begin{equation}
 D_\mu\,\theta=\partial_\mu\theta+2\,g_0\,\ma A^1_\mu\,.
\end{equation}
Note that $\rho$ and $\theta$ describe a gauged sigma-model with the 2d Euclidean hyperbolic target space. The scalar $\theta$ has a local shift symmetry and hence it is (locally) pure gauge, which implies it will be absent in the scalar potential
\eqref{scalarpotential}. We find that
\begin{equation}
\label{potential}
V=-4 \,g_0^2 \,e^{\xi_1+\xi_2}\cosh (\rho )+2\,
  g_0^2 \, e^{2 \xi_1} \sinh
   ^2(\rho )+\frac{k_0^2}{2} \,e^{2
   (\xi_1+ \xi_2)}\,.
\end{equation}
This can be derived from a superpotential $W$ given as 
\begin{equation}
\label{superpotential}
W=\frac{e^{\xi_2}}{2}\,\left(-2\,g_0+k_0\,e^{\xi_1}\right)-g_0\,e^{\xi_1}\,\cosh\rho
\, ,
\end{equation}
where $V=2[\sum_i (\partial_{\phi^i} W)^2 -W^2]$. We have checked that this truncation is consistent with the field equations of the full 3-dimensional theory.

As explained in the appendix \ref{newapp}, bosonic solutions of this truncated model preserve some supersymmetry if and only if the supersymmetry conditions \eqref{BPScond1} and \eqref{BPScond2} are satisfied, which for convenience we reproduce here:

\begin{equation}
 \begin{split}
 & 0 =\gamma^\mu\,\partial_\mu \xi_1\,\zeta_a-\gamma^\mu (\sqrt{-g})^{-1} \,{\varepsilon}_{\mu}^{\,\,\,\sigma\rho}\,\ma F^1_{\rho\sigma}\,\epsilon_{ab}\zeta^b+\left(k_0\,e^{\xi_1+\xi_2}-2\,g_0\,e^{\xi_1}\cosh(\rho)\right)\zeta_a\,,\\
 & 0 =\gamma^\mu\,\partial_\mu \xi_2\,\zeta_a-\gamma^\mu\,(\sqrt{-g})^{-1} {\varepsilon}_{\mu}^{\,\,\,\sigma\rho}\,\ma F^2_{\rho\sigma}\,\epsilon_{ab}\zeta^b+\left(k_0\,e^{\xi_1+\xi_2}-2\,g_0\,e^{\xi_2}\right)\zeta_a\,,\\
 & 0 =\gamma^\mu\,\partial_\mu\rho \,\zeta_a+\sinh(\rho)\gamma^\mu\,D_\mu \theta\,\epsilon_{ab}\zeta^b-2\,g_0\,e^{\xi_1}\sinh(\rho)\zeta_a\,,\\
  & 0 =\left(\partial_\mu + \frac{1}{4} \omega_{\mu}^{\, \, bc}\gamma_{bc}\right)
  \,\zeta_a+\frac{1}{4}\left(1-\cosh(\rho)\right)\,D_\mu \theta\,\epsilon_{ab}\zeta^b-2(\sqrt{-g})^{-1}\left(\,{\varepsilon}_{\mu}^{\,\,\,\sigma\rho}\,\ma F^1_{\rho\sigma}+\,{\varepsilon}_{\mu}^{\,\,\,\sigma\rho}\,\ma F^2_{\rho\sigma} \right)\epsilon_{ab}\zeta^b\\
 &+\left(\frac{g_0}{2}e^{\xi_2}-\frac{k_0}{4}e^{\xi_1+\xi_2}+\frac{g_0}{2}e^{\xi_1}\cosh(\rho)\right)\gamma_{\mu}\zeta_a\,.
 \label{BPS}
 \end{split}
\end{equation}
Note that we have recast the 4 real components of the supersymmetry parameter $\epsilon_i$ in a doublet of complex numbers $\zeta^a$ with $\zeta^1=\epsilon^1+i\epsilon^2$ and $\zeta^2=\epsilon^3+i\epsilon^4$.

Finally, let us comment on the vacua of the theory. From the supersymmetry conditions \eqref{BPS}, it directly follows that maximal supersymmetry is equivalent to
\begin{equation}
\rho=0 \, , \qquad \xi_1=\xi_2=\log\frac{2g_0}{k_0} \,
\end{equation}
where we chose $k_0$ and $g_0$ to be positive. One checks that indeed the potential is minimized for these values at
\begin{equation}
V=-\frac{8g_0^4}{k_0^2}\,.
\end{equation}
So, the maximally supersymmetric vacuum is AdS$_3$ as expected. More surprisingly this is only one of a family of AdS$_3$ solutions of the same curvature. This is due to a flat direction in the potential parameterized by $\xi_-=\xi_1-\xi_2$, since
\begin{equation}
\rho=0\,,\ \  \xi_1+\xi_2=2\log\frac{2g_0}{k_0}\quad \Rightarrow\quad  V=-\frac{8g_0^4}{k_0^2}\,.
\label{extrema}
\end{equation}
From the inspection of the first supersymmetry condition in \eqref{BPS} one concludes that whenever $\xi_-\neq 0$ these AdS$_3$ vacua break all supersymmetry. Note that from a 6d perspective, non-zero $\xi_-$ corresponds to a deformation of the S$^3$.
There are no other extrema of the potential \eqref{potential} other than \eqref{extrema}.

\section{Supersymmetric String Solutions in $D=3$}

Now, we would like to find supersymmetric string solutions in the $U(1)^2$ truncation 
\eqref{actionN=4model} with all vector fields and one of the scalar fields vanishing, i.e.
\begin{equation}
 \ma A^1_\mu=\ma A^2_\mu= \rho= 0 \,.
 \label{DWtruncation1}
\end{equation}
With this choice the scalar $\theta$ decouples from the BPS conditions \eqref{BPS} and the equations of motions. The 3d background describing a domain wall driven by the scalars $\xi_1$ and $\xi_2$ takes the form
\begin{equation}
 \begin{split}
  &ds^2_3=dr^2+e^{2U(r)}ds^2_{\mathbb{R}^{1,1}}\,,\\
  &\xi_1=\xi_1(r)\,,\\
   &\xi_2=\xi_2(r)\,.
  \label{DWansatzx1x2}
 \end{split}
\end{equation}
We consider a Killing spinor of the form
\begin{equation}
 \zeta_a(r)=Z(r)\,\zeta_{0\,a}\,,
 \label{spinorDW}
\end{equation}
with $\zeta_{0\,a}$ constant spinor and impose the condition
\begin{equation}
 \zeta_{0\,a}=\gamma^3 \zeta_{0\,a}\,,
 \label{proj}
\end{equation}
which breaks half of the supersymmetry. Here $\gamma^3$ is the Dirac matrix corresponding to the $r$-direction. We choose for the flat 3d Clifford algebra the following matrices
\begin{equation}
 \gamma^1=i\sigma^2\,,\qquad \gamma^2 =\sigma^3\,,\qquad \gamma^3=\sigma^1\,,
 \label{diracmatrices}
\end{equation}
where the Pauli spin matrices are
\begin{equation}
 \label{Pauli}
\sigma^{1} =
\left(
\begin{array}{cc}
0 & 1  \\
1 & 0
\end{array}
\right)
\hspace{5mm} \textrm{ , } \hspace{5mm}
\sigma^{2} =
\left(
\begin{array}{cc}
0 & -i  \\
i & 0
\end{array}
\right)
\hspace{5mm} \textrm{ , } \hspace{5mm}
\sigma^{3} =
\left(
\begin{array}{cc}
1 & 0  \\
0 & -1
\end{array}
\right) \ .
\end{equation}
Now the BPS equations \eqref{BPS} become 
\begin{equation}
\begin{split}
& U^\prime=-g_0\,e^{\xi_1}-g_0\,e^{\xi_2}+\frac{k_0}{2}\,e^{\xi_1+\xi_2}\,,\\
&\xi_1^\prime=2g_0\,e^{\xi_1}-k_0\,e^{\xi_1+\xi_2}\,,\\
&\xi_2^\prime=2g_0\,e^{\xi_2}-k_0\,e^{\xi_1+\xi_2}\,.\\
\label{BPSDWx1x2}
\end{split}
\end{equation}
The function $Z(r)$ in the Killing spinor \eqref{spinorDW} satisfies
\begin{equation}
Z^\prime=\frac12\,\left(-g_0\,e^{\xi_1}-g_0\,e^{\xi_2}+\frac{k_0}{2}\,e^{\xi_1+\xi_2}\right)Z\,,\\
\end{equation}
which, from \eqref{BPSDWx1x2} can immediately be solved as $Z=e^{U/2}$.

\subsection{Single Scalar Field}

It is clear that the above BPS equations \eqref{BPSDWx1x2} simplify drastically if the two scalars are equal, so we consider this case first. Let 
\begin{equation}
 \xi=\xi_1=\xi_2 \,.
 \label{DWtruncation2}
\end{equation}
Then by taking the scalar $\xi$ as the radial coordinate, the equation for $U$ 
in \eqref{BPSDWx1x2} is solved as
\begin{equation}
 e^{2U}=e^{-2\xi}\left(2g_0-k_0\,e^\xi\right)\,,
\end{equation}
where an integration constant is chosen as zero without loss of generality.
The 3d metric takes the form
\begin{equation}
 ds^2_3=\frac{e^{-2\xi}\,d\xi^2}{\left(2g_0-k_0e^\xi\right)^2}+e^{-2\xi}\left(2g_0-k_0e^\xi\right)ds^2_{\mathbb{R}^{1,1}}\,.
 \label{DWsolution}
\end{equation}
In the limit $\xi\rightarrow\log\left(2g_0/k_0 \right)$, the scalar curvature of \eqref{DWsolution} takes the negative constant value $-24g_0^4/k_0^2$, while for $\xi \rightarrow -\infty$ the scalar curvature vanishes. It is easy to see that
our domain wall solution interpolates between $\mrm{AdS}_3$ and a cone over $\mathbb{R}^{1,1}$. Note that if one crosses the horizon at $\xi\rightarrow\log\left(2g_0/k_0 \right)$ there is a signature change. We have checked that this solution satisfies the field equations of \eqref{so4action} too.

\subsection{Two Scalar Fields}
We now want to solve \eqref{BPSDWx1x2} for $\xi_1 \neq \xi_2$. If we define  
$X=e^{-\xi_1}$ and $Y=e^{-\xi_2}$, from the scalar field equations we get
\begin{equation}
Xe^{-\frac{2g_0}{k_0}X}= c_1 Ye^{-\frac{2g_0}{k_0}Y} \, .
\label{lambert}
\end{equation}
The constant $c_1$ has to be chosen as 1 so that we have a supersymmetric AdS$_3$ limit
which requires $X \rightarrow Y$ as we explained above\footnote{One may wonder if 
equation \eqref{lambert} with $c_1=1$ has any solution other than $X = Y$. Indeed, it has; 
the inverse of $f(x)= xe^x$ is the Lambert W function which has two real branches (see e.g. \cite{DBLP:journals/corr/abs-1209-0735}).}. 
Now introducing a new radial coordinate $R$ such that
\begin{equation}
 \frac{dR}{dr}=2g_0\ (e^{\xi_2} - e^{\xi_1}) \,,
\end{equation}
one finds that the solution of \eqref{BPSDWx1x2} is:
\begin{equation}
\begin{split}
 &e^{-\xi_1}=\frac{k_0}{2g_0}\,\frac{R\,e^{R}}{(e^R-1)}\,,\\
 &e^{-\xi_2}=\frac{k_0}{2g_0}\,\frac{R}{(e^R-1)}\,,\\
 &e^{2U}=\frac{k_0}{2g_0}\,\frac{R\,e^R}{(1-e^R)^2}\,.
 \label{sol2xi}
 \end{split}
\end{equation}
The 3d metric reads
\begin{equation}
 ds^2_3=\frac{k_0^2}{16\,g_0^4}\,\frac{R^2\,e^{2R}}{(1-e^R)^4}\,dR^2 +\frac{k_0}{2g_0}\,\frac{R\,e^R}{(1-e^R)^2}\,ds^2_{\mathbb{R}^{1,1}}\,.
 \label{sol2metric}
\end{equation}
It is straightforward to verify that this solution satisfies the field equations of our model \eqref{actionN=4model}.
In the limit $R\rightarrow 0$ the solution approaches to the $\mrm{AdS}_3$ vacuum of the model with the two scalars taking the value $e^{-\xi_1}=e^{-\xi_2}=k_0/2g_0$. The opposite limit $R\rightarrow +\infty$ is singular.

\section{Uplifts to $D=6$}
In \cite{Deger:2014ofa} it was shown that our 3-dimensional model given by the Lagrangian \eqref{so4action} can be
obtained from $D=6$ minimal supergravity coupled to a chiral tensor multiplet \eqref{Lag6} by a consistent S$^3$ 
compactification using the reduction ansatz found in \cite{Cvetic:2000dm}. When the gauge fields are zero, like in our solutions, this ansatz takes the form:
\begin{eqnarray}
ds_6^2&=&(\detT^{\frac{1}{4}})\left(\poped^{\frac{1}{2}}ds_3^2+\popeg^{-2}\poped^{-\frac{1}{2}}\Tmat^{-1}_{\ifoura\ifourb}
d\popem^\ifoura d\popem^\ifourb\right),\nonumber\\
\dil &=& \frac{1}{\sqrt{2}}\log\left(\poped^{-1}\detT^\frac{1}{2}\right)\label{Gansatz} , \\
\threeformsix &=&\gprime (\detT) \,\volthree-\frac{\poped^{-2}}{6g_0^2}\epsilon_{\ifoura\ifourb\ifourc\ifourd}\left(\tilde{U}\popem^\ifoura
d\popem^\ifourb\wedge d\popem^\ifourc\wedge d\popem^\ifourd 
+3d \popem^\ifoura\wedge d\popem^\ifourb\wedge d\Tmat_{\ifourc\ifoure}
\Tmat_{\ifourd\ifourf}\popem^\ifoure\popem^\ifourf \right) \, , \nonumber
\end{eqnarray}
where
\begin{eqnarray}
\popem^\ifoura\popem^\ifoura = 1\,,\qquad \poped=\Tmat_{\ifoura\ifourb}\popem^\ifoura\popem^\ifourb\,,
\qquad\tilde{U}=2\,\Tmat_{\ifoura\ifourc}\Tmat_{\ifourb\ifourc}\popem^\ifoura\popem^\ifourb
-\poped\Tmat_{\ifoura\ifoura}\, .
\end{eqnarray}
Now we will uplift the supersymmetric string solutions that we found in the previous section to $D=6$ with the help of this ansatz.
Since the compactification is consistent, they will automatically be supersymmetric solutions of the 6-dimensional theory.

\subsection{Uplift of the Single Scalar Solution}

In this case the scalar matrix \eqref{parT} takes the simple form
\begin{equation}
 T_{ij}=e^\xi\delta_{ij}\,,
\end{equation}
and the relevant quantities for the uplift \eqref{Gansatz} are 
\begin{equation}
\begin{split}
&\Delta=e^\xi\,,\qquad \tilde{U}=-2 \,e^{2\xi}\,.
\end{split}\label{quantity}
\end{equation}
Now using \eqref{Gansatz} on our 3-dimensional solution \eqref{DWsolution} we find:
\begin{equation}
 \begin{split}
  &ds_6^2=e^{\frac{3\xi}{2}}\,ds^2_3+g_0^{-2}\,e^{-\frac{\xi}{2}}\,ds^2_{S^3}\,,\\
  & ds^2_3=\frac{e^{-2\xi}\,d\xi^2}{\left(2g_0-k_0e^\xi\right)^2}+e^{-2\xi}\left(2g_0-k_0e^\xi\right)ds^2_{\mathbb{R}^{1,1}}\,,\\
  &H_{(3)}=k_0\,e^{4\xi}\,\mrm{vol}_{(3)}+\frac{1}{g_0^2}\,\mrm{vol}_{S^3}\,,\\
  &\dil=\frac{\xi}{\sqrt{2}}\,.
  \label{6dDWuplift}
 \end{split}
\end{equation}
If we now make the change of variable
\begin{equation}
  e^\xi=\frac{2\,g_0}{(k_0+g_0^2\,r^2)}\,,
  \label{r&xi}
\end{equation}
the solution \eqref{6dDWuplift} becomes 
\begin{equation}
 \begin{split}
  &ds^2=H_p^{-1/2}H_q^{-1/2}\,ds^2_{\mathbb{R}^{1,1}}+H_p^{1/2}H_q^{1/2}\,dr^2+H_p^{1/2}H_q^{1/2}r^2\,ds^2_{S^3}\,,\\
  &H_{(3)}=\frac{1}{g_0^2}\,\mrm{vol}_{S^3}-\mrm{vol}_{\mathbb{R}^{1,1}}\,\wedge\,d\,H_q^{-1}\,,\\
  &e^{-\sqrt2\dil}=H_qH_p^{-1}\,,
  \label{string}
 \end{split}
\end{equation}
where
\begin{equation}
 H_p=\frac{1}{g_0^2r^2}\,,\qquad  H_q=\frac{1}{2g_0}+\frac{k_0}{2g_0^3 r^2}\,.
\end{equation}
This is the ``dyonic'' string solution found\footnote{To be coherent with the conventions of \cite{Deger:2014ofa}, the 6d dilaton appearing in \cite{Duff:1996cf,Gueven:2003uw} has been rescaled as $\dil\rightarrow-\sqrt2\dil$.} in 
\cite{Duff:1995yh} (see also \cite{Duff:1996cf,Gueven:2003uw}), but without an additive constant in $H_p$. The solution is smooth everywhere \cite{Gueven:2003uw}. As $r\rightarrow 0$ the metric approaches 
to $\mrm{AdS}_3\times$ S$^3$ geometry, the dilaton becomes constant and only the magnetic charge survives.
Whereas, in the limit $r\rightarrow \infty$ we have a cone over 
S$^3 \times \mathbb{R}^{1,1}$, the dilaton goes to minus infinity and only the electric charge remains.
Note that unlike the solution found in \cite{Duff:1995yh} the solution is not asymptotically 
Minkowski (but conformally flat) due to the absence of an additive constant in $H_p$. This is a direct consequence of the reduction ansatz (\ref{Gansatz}). From (\ref{6dDWuplift}) it is easy to see that the breathing mode (i.e. the volume of $S^3$ \cite{Bremer:1998zp}) and the 6d dilaton are both determined in terms of the scalar field $\xi$ in such a way that when the sphere decompactifies, the dilaton diverges instead of going to a constant as in \cite{Duff:1995yh}.

\subsection{Uplift of the Two Scalar Solution}
Now let us consider the uplift of the two scalar domain wall solution \eqref{sol2xi}. In this case the scalar matrix \eqref{parT} has the form:
\begin{equation}
T=\left(
 \begin{array}{cc}
  e^{\xi_1}\,\mathbb{I}_2 & 0_2\\
  0_2 & e^{\xi_2}\,\mathbb{I}_2
 \end{array} \right)
 \end{equation}
If we now choose Hopf coordinates on S$^3$:
\begin{equation}
\vec{\mu}=\left(\sin\frac{\eta}{2}\cos\frac{\phi+\psi}{2},\sin\frac{\eta}{2}\sin\frac{\phi+\psi}{2},\cos\frac{\eta}{2}\cos\frac{\phi-\psi}{2},\cos\frac{\eta}{2}\sin\frac{\phi-\psi}{2}\right) \, ,
\end{equation}
applying \eqref{Gansatz} to \eqref{sol2xi} we find 
\begin{eqnarray}
  &&ds^2_6=\frac{2[\cos\eta+\coth (\frac{R}{2})]^{-1/2}}{g_0^{3/2}k_0^{1/2}R^{1/2}} 
  \left[\frac{2g_0^3}{k_0\,R} \left(2\cos\eta\sinh^2(\frac{R}{2})+\sinh R  \right)\,ds^2_3 +ds^2_{\tilde{S}^3}   \right] \, , \nonumber \\
  && ds^2_3=\frac{k_0^2}{16\,g_0^4}\,\frac{R^2\,e^{2R}}{(1-e^R)^4}\,dR^2 +\frac{k_0}{2g_0}\,\frac{R\,e^R}{(1-e^R)^2}\,ds^2_{\mathbb{R}^{1,1}}\,,\label{ugly}
\end{eqnarray}
where $ds^2_{\tilde{S}^3}$ is the metric of the squashed 3-sphere given by
\begin{equation}
\begin{split}
 &ds^2_{\tilde{S}^3}=a(R, \eta)\,(\sigma^1)^2+b(R,\eta)\,((\sigma^2)^2+(\sigma^3)^2)\,,\\
 &a(R,\eta)=\frac{(k_0\,R)}{(16\,g_0)}\,\frac{1+e^R+(e^R-1)\cos\eta}{e^R-1}\,,\\
 &b(R)=\frac{k_0\,R}{8\,g_0\,(e^R-1)}\,,
 \end{split}
\end{equation}
and the left-invariant 1-forms are given by
\begin{equation}
  \sigma^1=d\eta\, \, , \quad
  \sigma^2=\sin \left(\frac{\eta}{2}\right)\,(d\psi+d\phi)\, \, , \quad
   \sigma^3=\sin \left(\frac{\eta}{2}\right)\,(d\psi-d\phi)\,.
\end{equation}
The 6d dilaton is
\begin{equation}
 e^{-\sqrt2\,\varphi}= 4\,a(R,\eta)\,,\label{dil}
\end{equation}
and the 3-form is given by
\begin{equation}
\begin{split}
& H_{(3)}=\frac{16\,g_0^4\,e^{-2R}\,(e^R-1)^4}{k_0^3\,R^4}\,\text{vol}_{3}\\
 &-\frac{e^{-R}\,(e^R-1)^2\left(128+127\,\cosh R+127\,\cos\eta\,\sinh R\right)}{g_0^2\,\left(\sinh R+\cos\eta\,(\cosh R-1)\right)^2}\,\text{vol}_{\tilde{S}^3} \, .
 \end{split} \label{3form}
\end{equation}
In these coordinates the solution is not transparent. To get more insight,
it is useful to think of $\mathbb{R}^4$ as $\mathbb{C}^2$ with two complex coordinates $z$ and $w$ which we collectively denote as $\vec{u}$. Hopf-Spherical coordinates are defined as
\begin{equation}
z=r\sin\frac{\theta}{2}e^{i\alpha} \, , \qquad w=r\cos\frac{\theta}{2}e^{i\beta} \, , \qquad \alpha=\frac{\phi+\psi}{2}\, , \qquad \beta=\frac{\phi-\psi}{2}\, .
\label{hopf}
\end{equation}
It will also be useful later to introduce 
\begin{equation}
r_1=|z|=r\sin\frac{\theta}{2}\, , \qquad r_2=|w|=r\cos\frac{\theta}{2} \, .
\label{coords}
\end{equation}
Let us point out that $r_1$ and $\alpha$ provide polar coordinates in the $w=\,$constant planes, while $r_2$ and $\beta$ provide polar coordinates on the $z=\,$constant planes. We now perform the coordinate transformation
\begin{eqnarray}
e^R&=&\frac{1+g_0^2r^2\cos\theta+\sqrt{1+g_0^4r^4+2g_0^2r^2\cos\theta}}{g_0^2r^2(1+\cos\theta)}\, , \\
\cos\eta&=&\sqrt{1+g_0^4r^4+2g_0^2r^2\cos\theta}-g_0^2r^2 \, ,
\end{eqnarray}
after which the metric \eqref{ugly}, the dilaton \eqref{dil} and the 3-form \eqref{3form} take the form
\begin{eqnarray}
ds_6^2&=&(H_p H_q)^{-\frac{1}{2}}ds^2_{\mathbb{R}^{1,1}}+(H_p H_q)^{\frac{1}{2}}ds^2_{\mathbb{R}^4} \, , \nonumber\\
e^{-\sqrt{2}\dil} &=&H_qH_p^{-1} \, , \label{nice} \\
H_{(3)}&=&g_0^2r^4H_p^2\,\mrm{vol}_{\Omega^3}-\mrm{vol}_{\mathbb{R}^{1,1}}\,\wedge\,d\,H_q^{-1}\,, \nonumber
\end{eqnarray}
where in our new coordinates
\begin{equation}
ds^2_{\mathbb{R}_4}=dr^2+r^2 d\Omega_3^2 \, , \qquad d\Omega_3^2=\frac{1}{4}(d\theta^2+d\phi^2+d\psi^2-2\cos\theta d\phi d\psi)
\end{equation}
and
\begin{eqnarray}
H_p&=&\frac{1}{\sqrt{1+g_0^4r^4+2g_0^2r^2\cos\theta}} \, , \label{mag}\\
H_q&=&\frac{k_0}{2g_0}\log\frac{1+g_0^2r^2\cos\theta+\sqrt{1+g_0^4r^4+2g_0^2r^2\cos\theta}}{g_0^2r^2(1+\cos\theta)}\, .
\label{el}
\end{eqnarray}
One can verify that both $H_p$ and $H_q$ are indeed harmonic functions, i.e. solutions of the Laplace equation on $\mathbb{R}^4$ in these coordinates. It is also easy to check that as $ r \rightarrow \infty$ the geometry becomes $\mathrm{AdS}_3 \times$ S$^3$. 
The curvature scalar diverges as $r \rightarrow 0$ for $\theta=0$ and $\theta=\pi$. Another singularity occurs
as $r \rightarrow 1/g_0$ at $\theta=\pi$. These correspond to locations of the sources as we will see below. Note that except the form of the harmonic functions, the solution \eqref{nice} looks exactly the same as our previous
dyonic string solution \eqref{string}. But unlike before, it is not possible to remove magnetic strings from the system by setting $k_0=0$ since there is no additive constant in $H_p$. Let us also note that the harmonic function $H_p$ occurred 
before e.g. in \cite{Lunin:2002bj, Niehoff:2012wu} and corresponds to a uniform circular source. Meanwhile, $H_q$ being a logarithmic harmonic function suggests a 2-dimensional overall transverse space. These observations are further clarified 
in the next part.

\subsubsection{The Physical Interpretation}
We can get the physical interpretation of the solution \eqref{nice}
through a few observations. First note that a point source, i.e. a string fully localized in $\mathbb{R}^4$ (and with worldvolume along $\mathbb{R}^{1,1}$) corresponds to both magnetic and electric harmonic functions of the form
\begin{equation}
H_{\mathrm{point}}=a+\frac{b}{r^2} \, ,
\end{equation}
since
\begin{equation}
\nabla^2_{\mathbb{R}^4}\left(\frac{1}{r^2}\right) =\delta^4(\vec{u}) \, .
\end{equation}
This also implies that more generically, a string density (or smeared string configuration) $\sigma(\vec{u})$ will give rise to a harmonic solution of the form
\begin{equation}
H_\sigma(u)=\int d^4v \frac{ \sigma(v)}{|\vec{u}-\vec{v}|^2} \, .
\end{equation}
The question is then, can we find electric and magnetic string density's such that
\begin{eqnarray}
H_p&=&\int d^4v \frac{ \sigma_p(v)}{|\vec{u}-\vec{v}|^2} \, ,\\
H_q&=&\int d^4v \frac{ \sigma_q(v)}{|\vec{u}-\vec{v}|^2} \, .
\end{eqnarray}
The answer is yes, as we will now explain. It turns out that the magnetic strings are smeared along a ring in the 
$w=0$ plane (that is $\theta=\pi$) in \eqref{hopf}, of radius
\begin{equation}
r_0=\frac{1}{g_0} \, .
\end{equation}
More precisely the magnetic string density is
\begin{equation}
\sigma_p=\frac{r_0}{2\pi}\delta^2(w,\bar w)\delta(r-r_0) \, ,
\end{equation}
and this follows from the following computation
\begin{eqnarray}
H_p&=&\int d^4v \frac{ \sigma_p(v)}{|\vec{u}-\vec{v}|^2}\\
&=&\frac{1}{2\pi}\int_0^{2\pi} d\alpha \frac{r_0^2}{r_2^2+r_1^2+r_0^2-2r_0r_1\cos\alpha}\\
&=&\frac{r_0^2}{\sqrt{r_0^4+2r_0^2(r_2^2-r_1^2)+(r_1^2+r_2^2)^2}} \, .
\end{eqnarray}
The fact that this is identical to \eqref{mag} follows directly via \eqref{coords}.

Additionally one finds that the electric strings are smeared inside a disk of radius $r_0$ in the $w=0$ plane. The electric string density is
\begin{equation}
\sigma_q=\frac{k_0}{2g_0\pi}\delta^2(w,\bar w)\theta(r-r_0) \, ,
\end{equation}
where $\theta(r-r_0)$ is the Heaviside step function.
This is verified by direct computation
\begin{eqnarray}
H_q&=&\int d^4v \frac{ \sigma_q(v)}{|\vec{u}-\vec{v}|^2}\\
&=&\frac{k_0}{2g_0\pi}\int_0^{r_0} dr\int_0^{2\pi} d\alpha \frac{r}{r_2^2+r_1^2+r^2-2rr_1\cos\alpha}\\
&=&\frac{k_0}{2g_0}\int_0^{r_0} dr\frac{2r}{\sqrt{r_0^4+2r_0^2(r_2^2-r_1^2)+(r_1^2+r_2^2)^2}}\\
&=&\frac{k_0}{2g_0}\log\frac{r_2^2+r_0^2-r_1^2+\sqrt{r_0^4+2r_0^2(r_2^2-r_1^2)+(r_1^2+r_2^2)^2}}{2r_2^2} \, .
\end{eqnarray}
As before, one checks directly that this is identical to \eqref{el} by using \eqref{coords}.

In summary, we have discovered that the solution \eqref{sol2metric}, when lifted to 6 dimensions \eqref{nice}, corresponds to a rather peculiar configuration of strings. All strings have their world-volume along the $\mathbb{R}^{1,1}$ spanned by $(t,x)$. They are however spread out in the $w=0$ subplane of the $\mathbb{R}^4$ transverse space. In particular the electric strings are distributed with constant density inside a disc of radius $r_0$ in this plane, while the magnetic strings are distributed along the edge of this disc, a circle of radius $r_0$ in the same $w=0$ subplane. In this plane we thus have the simple 
picture: 

\begin{center}
\includegraphics[scale=0.5]{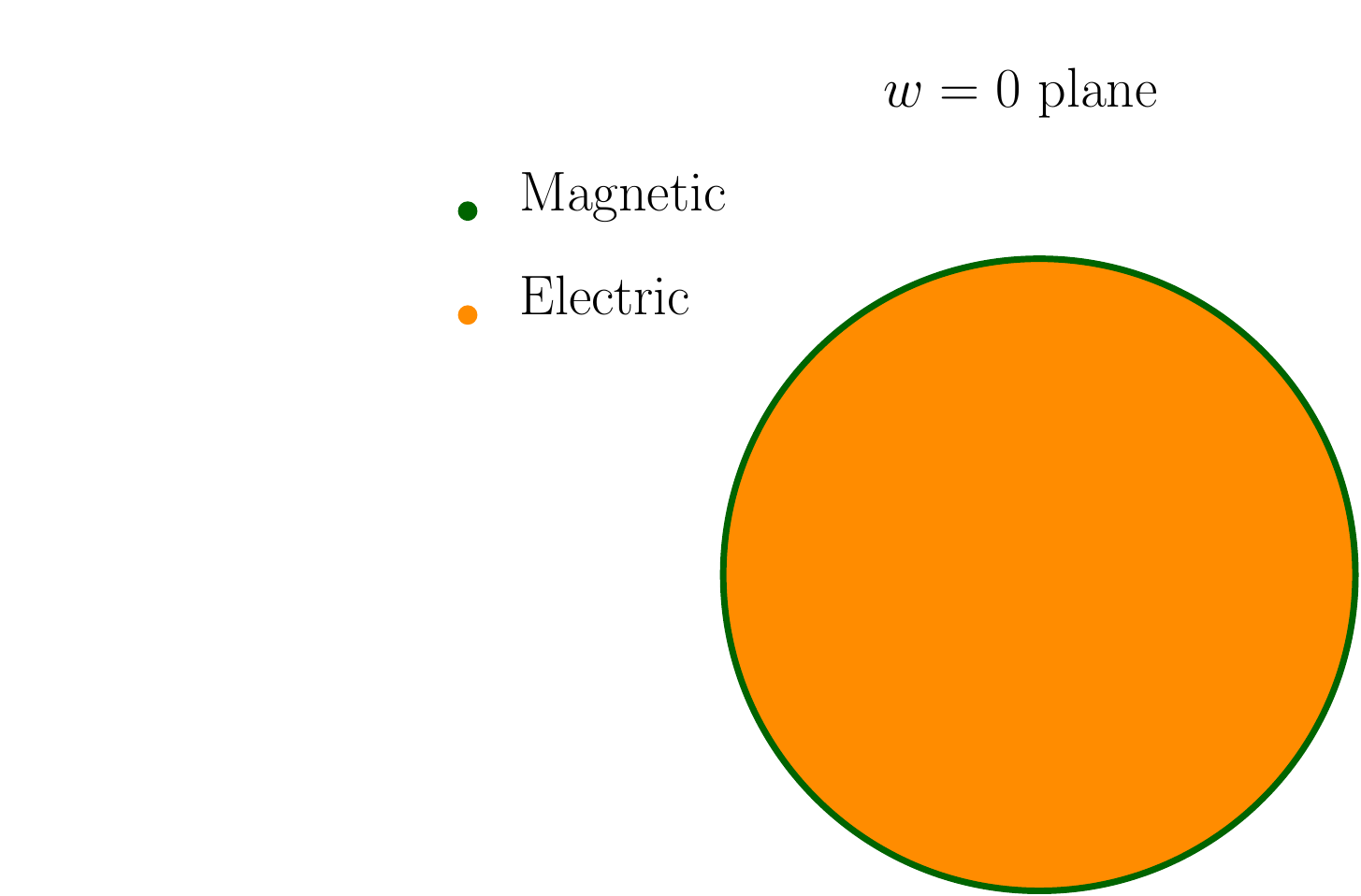}
\end{center}

It is well-established that in the case of a consistent sphere reduction, 
the higher dimensional origin of a domain wall is a brane distribution, see e.g. 
\cite{Bremer:1998zp, Kraus:1998hv, Freedman:1999gk, Cvetic:1999xx, Bakas:1999fa, Cvetic:2000eb, Cvetic:2000zu, Bergshoeff:2004nq}. 
But unlike ours, in most of the known examples these distributions are with non-dilatonic branes. Finally, we would like to point out that such BPS configurations of dyonic strings 
can directly be obtained studying 6d equations but only a small subset of them 
comes from our particular 3d gauged supergravity (\ref{so4action}).

\section{Final remarks}
\label{sec:final}
Consistent compactifications provide a powerful tool to obtain complicated solutions 
in a higher dimensional theory from a lower dimensional one. Following this idea,
in this paper we first found two supersymmetric black string solutions in the 3-dimensional $\ma N=4$, $SO(4)$ gauged supergravity and then embedded them to the 6-dimensional ungauged $\ma N=(1,0)$ supergravity, using the fact that these two models are connected by a consistent sphere reduction \cite{Deger:2014ofa}. Although, one of these solutions gave rise to an already known dyonic string \cite{Duff:1995yh}, from the other we obtained an interesting configuration which certainly deserves further investigation. First of all, it would be interesting to understand its connection with superstrata \cite{Bena:2017upb} or supertube \cite{Emparan:2001ux, Elvang:2003mj} type solutions. The fact that one of the 
harmonic functions in this solution already appeared in such set-ups (e.g. in 
\cite{Niehoff:2012wu}) hints a possible relationship.
Studying its 10-dimensional interpretation in terms of D1-D5 branes and its dimensional reduction to $D=5$ and comparison with black rings 
\cite{Elvang:2004rt} would be very illuminating. One may also consider its generalizations with more number of active scalar fields or non-zero gauge fields in 3-dimensions. It is possible to add pp-waves travelling along the worldvolumes of the strings making use of its null Killing vector too \cite{Deger:2004mw}. Studying RG flows using these string solutions \cite{Deger:2002hv} is another attractive direction which will give valuable information about the dual CFT.

Recently consistency of the reductions of $D=6$, $\ma N=(1,1)$ and $\ma N=(2,0)$ supergravities on $\mathrm{AdS}_3 \times$ S$^3$ was shown \cite{Hohm:2017wtr}. It would be very interesting to repeat our analysis for these cases too. The uplift of our single scalar solution (\ref{string}) is not asymptotically flat unlike the dyonic string solution found in  \cite{Duff:1995yh}. This suggests a possible generalization of the reduction ansatz (\ref{Gansatz}) which is worth investigating. We hope to come back to these problems soon.

\section*{Acknowledgements}
NP is fully and NSD is partially supported by the Scientific and Technological Research Council of Turkey (T\" ubitak) Grant No.116F137, DVdB is partially supported by Boğaziçi University Research Fund under grant number 17B03P1. NP and NSD are grateful to Abdus Salam ICTP where some parts of this paper were written. NSD also wishes to 
thank IHES for hospitality during the course of this work. We thank Can Koz\c{c}az for his collaboration at the beginning of this project. We thank Giuseppe Dibitetto, Roberto Emparan, Eoin \'O Colg\'ain and Tom\'as Ort\'in for useful discussions.

\appendix

\section{Supersymmetry equations}
\label{newapp}
Here we give derivation of the relevant supersymmetry equations for the theory \eqref{so4action}.

\subsection{From $SO(4)$ Yang-Mills to $\left(\mathbb{R}^3 \rtimes SO(3)\right)^2$ Chern-Simons} 
In 3-dimensions vectors are dual to scalars, which implies that one can always rewrite the theory in such a way that no dynamical\footnote{In this context 'dynamical' means 'with quadratic kinetic term'.} vectors are present, and only topological, i.e. Chern-Simons (CS), vectors remain. This CS formulation is the simplest and most natural setting in which to construct three dimensional gauged supergravity from the bottom-up \cite{deWit:2003ja, deWit:2004yr}. From the top-down perspective of dimensional reduction, one naturally ends up with dynamical vector fields, and one obtains the gauged supergravity in the so called Yang-Mills (YM) formulation. The precise connection and translation between these two formulations of three dimensional gauged supergravities was worked out in \cite{Nicolai:2003bp}. In that work a particular basis for the gauge group and embedding tensor were used which is slightly different from the one obtained from the sphere reduction in \cite{Deger:2014ofa}, which amounts to expressing $SO(4)$ as $SO(3)\times SO(3)$.

In summary, although our model \eqref{so4action} is a YM formulated theory based on the gauge group SO(4), to implement the results of \cite{deWit:2003ja, deWit:2004yr} one has to reformulate it as a CS theory based on the gauge group $\mathbb{R}^6\rtimes SO(4)\simeq\left(\mathbb{R}^3 \rtimes SO(3)\right)\times \left(\mathbb{R}^3 \rtimes SO(3)\right)$, via \cite{Nicolai:2003bp}. As most steps are rather straightforward (though somewhat tedious) applications of \cite{Nicolai:2003bp,deWit:2003ja,deWit:2004yr} we only present a few key formulae in this reformulation. 

The isomorphism between the adjoint representation of $SO(4)$ and $SO(3)\times SO(3)$ is realized by 't Hooft symbols $\eta^\alpha_{aij}$ \cite{tHooft:1976snw}. They take an antisymmetric pair of indices $[ij]$, $i,j=1,\dots, 4$ and map the (anti)self-dual component, in case $\alpha=(+, -)$ respectively, to the index $a=1,2,3$. Explicitly

\begin{equation}
 \eta^\pm_{aij}=\epsilon_{aij}\pm \delta_{ai}\delta_{j4} \mp \delta_{aj}\delta_{4i}\,.
\end{equation}
This change of indices on any tensor $V$ is then implemented in practice via the formulae
\begin{equation}
V_{ij}=\eta_{aij}^\alpha V_a^\alpha\, , \qquad V_a^\alpha=\frac{1}{4}\eta_{aij}^\alpha V_{ij}
\end{equation}
In the CS formulation one replaces the dynamical degrees of freedom in the field strengths $F$ of \eqref{so4action} by additional scalars $\chi$ and introduces extra (topological) vectors, $G=dC$, that gauge an extra nilpotent factor in an enlarged gauge group \cite{Nicolai:2003bp}. The precise relations between the CS and YM fields are given by the duality formulae
\begin{equation}
\begin{split}
&2\, \sqrt{-g}\,\varepsilon_{\mu\nu\rho}\,D^{\rho}\chi^\alpha_a=-M_{ab}^{\alpha\beta}\,F_{\mu\nu\,b}^\beta\,,\qquad  M^{\alpha\beta}_{ab}=T^{-1}_{ik}T^{-1}_{jl} \eta^\alpha_{aij}\eta^\beta_{bkl}\,,\\
&G^{\pm}_{\mu\nu\, a}-\frac12\,\epsilon_{abc}\,\chi_b^\pm F^\pm_{\mu\nu\,c}=\frac14\,\sqrt{-g}\, \varepsilon_{\mu\nu\rho}\,\eta_{a\,jk}^\pm\,T^{-1}_{ij}\,D_\mu T_{ik}\,.
\label{constraints}
\end{split}
\end{equation}

\subsection{The Supersymmetry Variations}
\label{SUSYvariations}
Once the connection to the CS formulation has been made, one can obtain the supersymmetry variations for the $\ma N=4$ gauged supergravity associated to \eqref{so4action} from \cite{deWit:2003ja}. As our interest is in studying bosonic solutions preserving supersymmetry, we only present the fermionic variations and assume that all fermionic fields vanish. The scalar fields of the CS formulation, $\varphi$ in \cite{deWit:2003ja}, are conveniently grouped as
\begin{equation}
U_{ij}= T_{ij} + \chi_{ij} \, .
\end{equation}
The supersymmetry variations of \cite{deWit:2003ja} involve a number of geometric data of the scalar manifold and details on the embedding of the gauge group into the isometry group of that scalar manifold. For our 3d model these were worked out in \cite{Deger:2014ofa}. Combining these results leads to
\begin{eqnarray}
  \delta_\epsilon\,\psi_\mu^i&=&\hat{\nabla}_\mu\,\epsilon^i-\frac12\, W\,\gamma_\mu\,\epsilon^i\,,\label{YMSUSYN=4gravitino}\\
  \delta_\epsilon\,\lambda^{i\,,kl}&=&\frac12\,\left(\gamma^\mu\,D_\mu\,U^{mn}+\partial^{mn} W\right)\left(\delta^{ij}\delta^k_{m}\delta^l_n-f^{ij\,\,kl}_{\quad mn}\right)\,\epsilon_j\,,\label{variationchi}
\end{eqnarray}
where
\begin{equation}
 \begin{split}
  &\hat{\nabla}_\mu\,\epsilon^i=\nabla_\mu\,\epsilon^i+\mathbb{P}_{+}^{\,ij kl}\,\epsilon_j\,\left( \sqrt{T^{-1}}D_\mu \sqrt{T}\right)_{kl}+\mathbb{P}_{+}^{\,ij kl}\,\epsilon_j\sqrt{T}_{km}\,D_\mu\chi_{mn}\,\sqrt{T}_{nl}\,,
  \label{modifiedcovdevCS}
 \end{split}
\end{equation}
where $\nabla_{\mu}= (\partial_\mu + \frac{1}{4} \omega_{\mu}^{\, \, bc}\gamma_{bc})$,
$\mathbb{P}^+_{ijkl}=\frac{1}{4}\left(\delta_{ik}\delta_{jl}-\delta_{jk}\delta_{il} + \epsilon_{ijkl}   \right)$ and the superpotential is
 \begin{equation}
 W=\frac{1}{2}\,\left( k_0\,\det\sqrt T-g_0\,\text{Tr}\,T \right)\,.
 \label{superpot}
\end{equation}
Finally, the complex structures $f^{ij}$ on the scalar target space are 
\begin{equation}
f^{ij}=-(\Gamma^{ij})_{kl\,mn}\,e^{kl}\wedge e^{mn}\,,\qquad \text{with}\qquad (\Gamma^{ij})^{kl}_{\,\,\,\,mn}=4\,\delta^{km}\,\mathbb{P}^{ijln}_+\,,
\end{equation}
where  $e_{mn}$ is the vielbein on the target manifold as in (5.2) of \cite{deWit:2003ja}.

\subsection{Supersymmetry Conditions for the Truncated Model}
We now investigate the vanishing of \eqref{variationchi} under the assumption that the only non-trivial bosonic fields are those of the truncation (\ref{parT}) and(\ref{parA}). Note that in the CS formulation the ${\cal F}^{1,2}$ of the main text 
\eqref{parA}
are related to scalars $\chi^{1,2}$ through the duality relations \eqref{constraints}, which after the truncation become:
\begin{equation}
e^{\xi_{1,2}}D_\mu \chi^{1,2}= \frac{\varepsilon_\mu{}^{\nu\rho}{\cal F}_{\nu\rho}^{1,2}}{\sqrt{-g}} \, .
\end{equation}
 Labeling the scalars appearing in this truncation 
as $\phi^i=(\xi_1,\xi_2,\rho,\theta)$ and $\chi^I=(\chi^1,\chi^2)$, we can formally re-express the vanishing of \eqref{variationchi} as
\begin{equation}
\left(\frac{\partial U^{mn}}{\partial \phi^p}\left(\gamma^\mu\,D_\mu\phi^p\right)\,+\frac{\partial U^{mn}}{\partial \chi^I}\left(\gamma^\mu\,D_\mu\chi^I\right)\,+\partial^{mn} W\right)\left(\delta^{ij}\delta^k_{m}\delta^l_n-f^{ij,\,\,kl}_{\quad mn}\right)\,\epsilon_j=0\,.\label{BPSstep1}
\end{equation}
Our approach to analyze these equations is to think of them as a linear algebra problem determining the variables $X_j^i=(D_\mu\phi^i)\gamma^\mu \epsilon_j$, since the equations \eqref{BPSstep1} have the form
\begin{equation}
M^{ikl\,j}{}_p X_j^p=V^{ikl}\,.\label{linalg}
\end{equation}
Note that these are 64 equations for 16 variables, and if $M$ and $V$ would be generic these equations would be without solution. However, and this might have been expected since the truncation (\ref{parT})-(\ref{parA}) is consistent, in this particular case most of the equations are actually redundant. It turns out that only 12 of them are linearly independent and so the 16 components of $X_j^i$ are not uniquely determined. We find it convenient to choose the 12 variables to solve for as $X_i^\alpha$, $\alpha=1,2,3$ and reorganize them as 6 complex variables $Z_a^\alpha$, $a=1,2$, defined as $Z_1^\alpha=X_1^\alpha+iX_2^\alpha$ and $Z_2^\alpha=X_3^\alpha+iX_4^\alpha$. Carrying through the straightforward but somewhat tedious solution of \eqref{linalg} one finds
\begin{equation}
 \begin{split}
 &Z_a^1= \gamma^\mu\,\partial_\mu \xi_1\,\zeta_a =\gamma^\mu (\sqrt{-g})^{-1} \,{\varepsilon}_{\mu}^{\,\,\,\sigma\rho}\,\ma F^1_{\rho\sigma}\,\epsilon_{ab}\zeta^b-\left(k_0\,e^{\xi_1+\xi_2}-2\,g_0\,e^{\xi_1}\cosh(\rho)\right)\zeta_a\,,\\
 &Z_a^2= \gamma^\mu\,\partial_\mu \xi_2\,\zeta_a =\gamma^\mu\,(\sqrt{-g})^{-1} {\varepsilon}_{\mu}^{\,\,\,\sigma\rho}\,\ma F^2_{\rho\sigma}\,\epsilon_{ab}\zeta^b-\left(k_0\,e^{\xi_1+\xi_2}-2\,g_0\,e^{\xi_2}\right)\zeta_a\,,\\
 &Z_a^3= \gamma^\mu\,\partial_\mu\rho \,\zeta_a=-\sinh(\rho)\gamma^\mu\,D_\mu \theta\,\epsilon_{ab}\zeta^b-2\,g_0\,e^{\xi_1}\sinh(\rho)\zeta_a\,.
 \label{BPScond1}
 \end{split}
\end{equation}
with $\zeta^1=\epsilon^1+i\epsilon^2$ and $\zeta^2=\epsilon^3+i\epsilon^4$.

Additionally to the vanishing of \eqref{variationchi} one also needs to impose the vanishing of the gravitino variation \eqref{YMSUSYN=4gravitino}. Inserting the truncation (\ref{parT}) and (\ref{parA}) into \eqref{YMSUSYN=4gravitino} gives
\begin{eqnarray}
0&=&\nabla_\mu\,\zeta_a+\frac{1}{4}\left(1-\cosh(\rho)\right)\,D_\mu \theta\,\epsilon_{ab}\zeta^b-2(\sqrt{-g})^{-1}\left(\,{\varepsilon}_{\mu}^{\,\,\,\sigma\rho}\,\ma F^1_{\rho\sigma}+\,{\varepsilon}_{\mu}^{\,\,\,\sigma\rho}\,\ma F^2_{\rho\sigma} \right)\epsilon_{ab}\zeta^b\nonumber\\
&&+\left(\frac{g_0}{2}e^{\xi_2}-\frac{k_0}{4}e^{\xi_1+\xi_2}+\frac{g_0}{2}e^{\xi_1}\cosh(\rho)\right)\gamma_{\mu}\zeta_a\,. \label{BPScond2}
\end{eqnarray}
To summarize, any bosonic solution of the $U(1)^2$ truncated model \eqref{actionN=4model} 
that preserves some supersymmetry should satisfy \eqref{BPScond1} and \eqref{BPScond2}.

\bibliographystyle{utphys}
\bibliography{references}
\end{document}